\documentclass[prl,twocolumn,superscriptaddress,showpacs]{revtex4}

\def\beq{\begin{equation}}
\def\eeq{\end{equation}}
\def\beqn{\begin{eqnarray}}
\def\eeqn{\end{eqnarray}}
\usepackage{graphicx}
\begin{document}
\title{Quantum destruction of stiffness in diluted antiferromagnets and superconductors}

\author{N. Bray-Ali}
\affiliation{Department of Physics, University of California, Berkeley, CA 94720}
\author{J. E. Moore}
\affiliation{Department of Physics, University of California, Berkeley, CA 94720}
\affiliation{Materials Sciences Division, Lawrence Berkeley National Laboratory, Berkeley, CA 94720}
\date{\today}

\begin{abstract}
The reduction of 2D superconducting or antiferromagnetic order by
random dilution is studied as a model for the 2D diluted Heisenberg
antiferromagnet (DHAF) La$_2$Cu$_{1-p}$(Zn,Mg)$_p$O$_4$ and randomly inhomogeneous 2D suerconductors.   We show in simplified models that long-range order can persist at the percolation threshold despite the presence of disordered one-dimensional segments, contrary to the classical case.  When long-range order persists to the percolation threshold, charging
effects (in the superconductor) or frustrating interactions (in
the antiferromagnet) can dramatically modify the stiffness of the
order.  This quantum destruction of stiffness is used to model neutron scattering data on La$_2$Cu$_{1-p}$(Zn,Mg)$_p$O$_4$.  In a certain simplified model, there is a sharp stiffness transition between ``stiff'' and ``floppy'' ordered phases.
\end{abstract}
\pacs{75.10.Jm, 75.10.Nr, 75.40.Cx, 75.40.Mg}\maketitle
Randomly diluted superconductors and antiferromagnets involve a
combination of classical percolation physics with the quantum physics
underlying superconductivity and antiferromagnetism.  Percolation is
perhaps the simplest transition that can occur in a disordered system:
random dilution of sites (or bonds) on a lattice induces a transition
between a phase with one infinite nearest-neighbor connected cluster
of occupied sites (bonds), and a phase with only disconnected finite
clusters.  Recent experiments on diluted 2D antiferromagnets and
inhomogeneous superconductors require a theory of how
quantum-mechanical effects modify the percolation transition in these
systems.  This question is also of practical importance for
field-effect devices in which a thin film is tuned through the
superconducting transition~\cite{ahn}.

\begin{figure}
\includegraphics[width=3.0in]{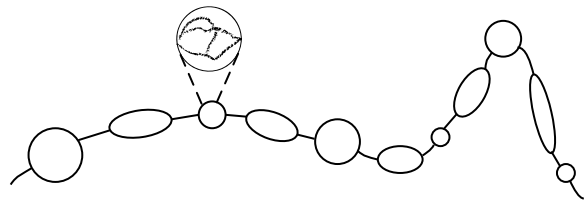}
\caption{The backbone of the incipient infinite cluster (the backbone
is the portion that carries current from one end of the sample to
another) showing 2D ``blobs'' and 1D ``links''.}
\label{figone}
\end{figure}

At the percolation threshold, $p_c$, the infinite connected cluster
is on the verge of being cut into disconnected finite clusters~(Fig.\ref{figone}).  For the
connected cluster to have long-range order (LRO), the quasi-2D ``blobs'' must
correlate across quasi-1D ``links.'' This occurs easily in some other
diluted magnets~\cite{senthil,coniglio} and inhomogeneous
superconductors~\cite{deutscher}, since their degrees of freedom have
LRO even in a 1D chain.  The two cases considered here, the $s=1/2$ Heisenberg antiferromagnet and the $O(N)$ quantum rotor, both have quantum degrees of
freedom that order in 2D but not in 1D.  Then the question of whether LRO survives to $p_c$, when the infinite cluster is a fractal object of dimension $\frac{91}{48}\approx 1.896$~\cite{bunde}, is unanswered.

The effect of quantum fluctuations is strongest on the quasi-1D chains.  For classical models on the percolation cluster at $T>0$, the fact that these chains are disordered (have a finite correlation length $\zeta$) implies that the cluster has no LRO.  Similar arguments have been made for the quantum cases discussed here~\cite{senthil}.  The first part of this paper shows that the existence of arbitrarily long 1D segments at $p_c$, and the fact that a spin or rotor in the middle of such a segment has strong quantum fluctuations, {\it does not} prevent LRO for the cluster: the 2D blobs can order through disordered 1D links, in a manner that is impossible for classical models.  These one-dimensional segments cause difficulty for spin-wave calculations because, as seen below, the spins in the middle of such a segment fluctuate strongly.  An additional motivation for studying the effect of these 1D segments is that their effects are essentially unobservable in current QMC studies, as a typical realization of e.g. 10$^5$ spins will contain no 1D segments of length longer than 8.  A different way of connecting 1D physics to diluted antiferromagnets is discussed in~\cite{eggert}.

Randomly diluted quantum degrees of freedom appear in two
well-known nanoscale inhomogeneous materials.
La$_2$Cu$_{1-p}$(Zn,Mg)$_p$O$_4$ is obtained by adding ``static holes'' (i.e., removing spins)
at random in a quasi-2D antiferromagnet.  For small hole
densities~$p\ll p_c$, neutron scattering measurements~\cite{vajk} agree well with quantum Monte Carlo \cite{vajk,sandvik,todo} and
spin-wave \cite{castroneto} calculations using the
$s=\frac{1}{2}$ site-diluted Heisenberg antiferromagnet (DHAF) model
\beq H = J \sum_{\langle i j \rangle} p_i p_j {\bf s}_i \cdot {\bf
s}_j, \label{dhaf} \eeq
where the sites form a square lattice, and $p_i,p_j$ equal 0 with
probability $p$ and 1 with probability $1-p$.  (Vajk
et.al.~\cite{bilayer} argue that other interactions besides those
in~(\ref{dhaf}) may not be neglected close to threshold.  We consider
some such interactions below.)

The second type of material, inhomogeneous high-temperature
superconductors like~Bi$_2$Sr$_2$CaCu$_2$O$_{8+\delta}$ (BSCCO), comes
from doping mobile holes into a quasi-2D antiferromagnet.
Scanning tunneling spectroscopy of BSCCO surfaces show grains of
size~$\approx 3$~nm that are either superconducting or
insulating~\cite{lang}.  The grains are larger than the coherence
length $\approx 1$~nm, so it is reasonable to assume that they contain
Cooper pairs, and that collectively they resemble a bond-diluted
Josephson-junction (JJ) model
\beq H_{JJ} = \sum_i E_C (n_i - n_0)^2 - E_J \sum_{\langle ij
\rangle} p_{ij} \cos(\theta_i - \theta_j). \label{jja}
 \eeq
Here the bond variables $p_{ij}$ have the same distribution as $p_i$ in~(\ref{dhaf}).  The above is not expected to be as accurate a model of the microscopics as the diluted Heisenberg model (\ref{dhaf}) is for La$_2$Cu$_{1-p}$(Zn,Mg)$_p$O$_4$, since the microscopic origin of the disorder in BSCCO is unknown.

The model (\ref{jja}) is in the universality class of the $O(2)$ rotor model, and for $E_C\neq0$, charging effects in the grains prevent LRO in a 1D chain~\cite{sachdev}.  The local charge~$n_i$ does not commute with the superconducting phase~$\phi_i$: charging tends to fix the local
{\it number}, which is conjugate to the local {\it phase}.  In the
zero-charging-energy limit~\cite{deutscher}, $H_{JJ}$ orders in
1D at $T=0$.  In the Heisenberg spin-half case, and for small nonzero $E_C/E_J$ in the Josephson-junction case, the 1D model is critical (has power-law correlations); for large $E_C/E_J$ or Heisenberg spin-one, the 1D model is short-ranged.


One of our main conclusions for both the Heisenberg antiferromagnet and $O(N)$ rotor is that order at $p_c$ is allowed despite the existence of 1D links: in a simplified limit we find a ``renormalized classical'' phase that provides an adequate description of some previous numerical results.  We find that possible ordered states at $p_c$ must have extremely low stiffness relative to the
undiluted 2D case.  At $T=0$, order and stiffness are independent quantities,
but mix as $T$ increases: low stiffness leads to a reduced $T>0$ correlation length $\lambda$.  Such order with low stiffness can be proven to occur if there are classical magnetic moments or superconducting phases (see (\ref{toy2})) coupled by
quantum 1D links.  Superconducting phases with small superfluid density have previously been proposed, e.g., the ``gossamer superconductor'' of~\cite{laughlin}.  Our picture differs in that the
low superfluid density results from randomness on scales larger than the coherence length~\cite{lang,trivedi}, rather than from a uniform theory.

\begin{figure}
\includegraphics[width=2.5 in]{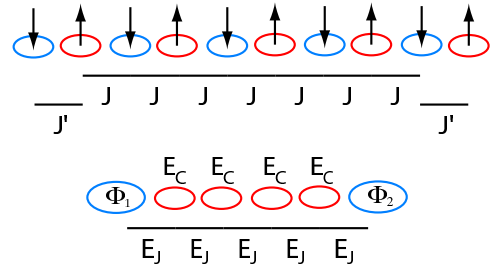}
\caption{Two examples of how quantum systems can order through a
disordered 1D segment.  The first case is an AF spin-half chain with
two weakly connected spins at the end; the second case is two bulk
superconductors ($E_C=0$) connected by a chain of grains with nonzero
$E_C$.}
\label{figtwo}
\end{figure}

We first calculate correlations in a toy model~(\ref{jprime}) to show explicitly how
quantum disordered or critical $T=0$ 1D systems like the spin-half chain are fundamentally different from classical disordered $T>0$ 1D systems.  Even though both may be disordered in 1D, the quantum systems can order through disordered 1D regions~(Fig.\ref{figtwo}) while similar classical systems cannot.  This simple statement underlies the possible existence of order at $p_c$:.  We next solve a rather simplified limit of (\ref{jja}) that shows a stiffness transition
between two ordered phases: a ``stiff'' renormalized classical phase,
and a ``floppy'' phase dominated by fluctuations on the 1D
links.  In a less simplified but still approximate model, the ``stiff phase'' becomes just a renormalized classical phase like in the 2D Heisenberg antiferromagnet (AF)~\cite{chakravarty}, except that the renormalized problem is not the 2D classical AF but the classical AF at $p_c$. Finally, we compare experiment and simulations of~(\ref{dhaf}) to this renormalized classical theory.

%
%

First consider a chain of spins with uniform coupling $J$, plus one spin at each end attached by a coupling $J^\prime$:
\begin{equation}
H_1 = \sum_{i=1}^{N-1} J {\bf s}_i \cdot {\bf s}_{i+1}+
J^{\prime}({\bf s}_0 \cdot {\bf s}_1+{\bf s}_N \cdot {\bf
s}_{N+1}). \label{jprime}
\end{equation}
Here, ${\bf s}_i$ are $s=\frac{1}{2}$ Heisenberg spins coupled
antiferromagnetically, $J\gg J^{\prime}>0$.  For very small
$J^\prime$, the state of the internal $N$ spins is nearly
undisturbed, and in particular ${\bf s}_1$ and ${\bf s}_N$ are
only weakly correlated with each other~\cite{eggert}.
 However, the two spins at the ends ${\bf s}_0$ and ${\bf s}_{N+1}$
can be made to form a perfect singlet with each other, for $N$
even: $\lim_{J^\prime /J \rightarrow 0^+} {\bf s}_0 \cdot {\bf
s}_{N+1} = -3/4.$

To understand this result, note that at $J^\prime = 0$, the four
states of the end spins, which can be classified into a singlet and
triplet, are degenerate.  Perturbation theory in $J^\prime$ 
preserves rotational symmetry and, for $J^\prime\ll J$, just
splits the singlet and triplet.  The two spins at the ends
correlate perfectly with each other, even though the internal
spins are weakly correlated.  Diagonalization, using
thick-restarted Lanczos~\cite{lanczos}, up to $N=20$ confirms the second-order perturbation theory argument.  At small enough $J^\prime/J$ there is an arbitrarily strong
singlet correlation
\beq
\langle s_A \cdot s_B \rangle \approx -\frac{3}{4} + f_1(N) \left({J^\prime \over J}\right)^2
\eeq
but with a vanishing energy splitting between singlet and triplet:
\beq
E_{\rm triplet} - E_{\rm singlet} \approx f_2(N) J \left({J^\prime \over J}\right)^2.
\eeq
Here $f_1$ and $f_2$ are dimensionless functions of the chain length that incorporate the matrix elements and energy denominators of the unperturbed chain.

At $p_c$, the Josephson junction array (\ref{jja}), in a certain
approximation, also exhibits long-range order.  Consider
(for now) neglecting all fluctuations within 2D blobs.   The scaling properties of this model can be found exactly; later, fluctuations within the blobs will be partially restored.  In this approximation,
the phases of the blobs, $\{\phi_i\}$, commute with $H$,
so choose a basis of simultaneous eigenstates of~(\ref{jja})
and $\phi_i$.  In this basis, the energy 
\begin{equation}
E\approx -\sum_{i}J_i\cos(\phi_i-\phi_{i+1}),
\label{toy2}
\end{equation}
depends on exchange constants $J_i>0$ chosen from a distribution
$P(J)$ which we calculate below using properties of the 1D links
(see~(\ref{distribution})).  Regardless of the distribution, we
have long-range order, since $J_i>0$ and $\phi_i$ are classical.
We calculate the stiffness of this order below using the $J\rightarrow0$ behavior of $P(J)$.



Consider a single 1D link of $L$ sites bounded by two superconducting
blobs of phase $\phi_1$ and $\phi_2$.  After the quantum-classical
mapping, we obtain a 2D XY model with fixed boundary
conditions~\cite{sachdev}.  The lowest energy is obtained if $\phi_1 =
\phi_2$, but the energy cost to create a phase difference depends
on the coupling $K=E_C/E_J$ in the 2D XY model.  This problem is
discussed in~\cite{cardy}: the stiffness per site goes to a finite
limit in the algebraic (ALRO) phase of the 2D XY model but falls off
exponentially with $L$ in the short-ranged phase (SRO).  For small
phase difference between boundaries $\Delta \phi = \phi_1 -
\phi_2$,
\beq E(\Delta\phi) \sim \cases{k_1 (\Delta \phi)^2 L^{-1}&in
ALRO phase\cr k_2 (\Delta \phi)^2 e^{-L/\zeta}&in SRO phase}.
\label{single}
\eeq
%
The exchange strength is $J(L) = 2 E(\Delta \phi) / (\Delta \phi)^2$.

The distribution of exchange strengths between blobs, $P(J)$, is chosen to reflect the geometry of the percolation cluster~\cite{coniglio,herrmann}.
The total number of 1D links between $A$ and $B$ goes as $N\sim R^{3/4}$
and the fraction of links $P(L)$ of length $L$ falls off
exponentially with $L$: $P(L) \sim e^{-a L}$.  

The probability distribution $P(J)$ of exchange strengths is a sum of $\delta$-functions:
\begin{equation}
P(J) = \sum_{L=1}^{N \sim R^{3/4}} P(L) \delta(J - J(L))
\end{equation}
The peaks become closely spaced as $J \rightarrow 0$.
Using~(\ref{single}) to change variables, these properties imply that
as $J\rightarrow0$, the fraction of links with exchange $J$ is
\begin{equation}
P(J \rightarrow 0) = \cases{ k_1/J^2 e^{-k_1a/J}&in ALRO phase\cr
(J/k_2)^{a\zeta-1}&in SRO phase}. \label{distribution}
\end{equation}
Any given link must be shorter than the total number of
links $N\sim R^{3/4}$, so the distribution~(\ref{distribution}) is cut off at
\begin{equation}
J_0=\cases{k_1/N &in ALRO phase\cr k_2e^{-N/\zeta}&in SRO phase}.
\label{cut}
\end{equation}

Now consider the total energy cost $E(R,\Delta \phi) = J_{\rm eff} (\Delta \phi)^2/2$ to create a small phase difference $\Delta \phi$ between two faraway points $A$ and $B$ separated by distance $R$, where $J_{\rm eff}$ is the disorder-averaged stiffness.  For the classical $XY$ chain with a random distribution of couplings $P(J)$, it is known~\cite{straley} that there is a transition depending on the exponent $P(J) \propto J^\alpha$ as $J \rightarrow 0$.  For $\alpha \geq 0$, there is effectively a nonzero mean stiffness, and $J_{\rm eff}(R) \sim R^{-3/4}$.  For $\alpha < 0$, there is a qualitatively weaker stiffness with exponent depending on $\alpha$.

Hence our model has a transition when $a = \zeta^{-1}$, or
when the typical length of a 1D link on the percolation backbone is
equal to the correlation length of the 1D JJ.  There are
estimates of the number $a$ for some standard
lattices~\cite{herrmann}.  The stiffness transition is {\it not} at
the same value of $E_J/E_C$ as the KT transition where the
1D correlation length becomes finite.  The transition is caused
by competition between the 1D correlation length $\zeta$ and the percolation
physics that controls
$a$.

To improve the model, the 2D blobs can be modeled as 2D classical percolation clusters rather than single moments.  (This is still an approximate model, since it neglects quantum effects in the blobs, of course.)  With this modification, the asymptotic behavior of the sum is dominated by the blobs in the stiff phase: now $E_\pi(R) \sim R^{-t/\nu}$ in the stiff phase, where $t \approx 1.31$ is the classical percolation resistivity exponent and $\nu = 4/3$.  The location of the transition to the floppy phase is also modified.  The conclusion is that, at least in this approximate model, there is a stiff phase  described by $E_\pi(R) \sim R^{-t/\nu}$ but with only ALRO, or even SRO with sufficiently large $\zeta$, on the 1D links.  Quantum fluctuations on the links do reduce the stiffness numerically in this stable ``stiff'' phase, and in the toy model can induce an exotic ``floppy'' phase.


In the physical system where the blobs also are quantum-mechanical, the sharp transition
between stiff and floppy phases will be smeared out since there is
no obvious order parameter to protect it; and there will appear a
disordered phase for large $E_C$.  This model may explain the numerical observation of~\cite{sandvik} on the spin-half Heisenberg model that the stiffness has the percolation exponent but is numerically less than the classical value.

The above description of the percolation cluster depended only on the assumption of either algebraic order or a correlation length in the one-dimensional links, so the same physical picture describes also the $O(N)$ rotor models for $N>2$ and the Heisenberg quantum antiferromagnet.  For the antiferromagnet, the 1D links are algebraically ordered for half-integer spin with short-ranged interactions, but develop a correlation length for either integer spin or strong frustrating interactions.

\begin{figure}
\includegraphics[width=2.8 in]{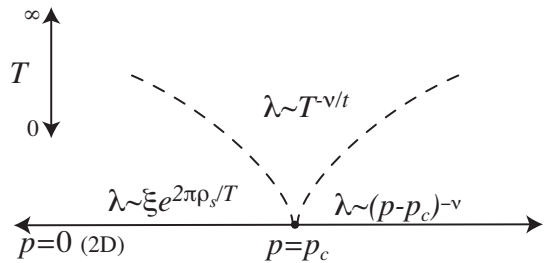}
\caption{The three different stiff-phase forms for the correlation
length $\lambda$ near the $T=0$, $p=p_c$ geometric critical point.
Above $p_c$ the correlation length is just set by the percolation
correlation length (cluster size) $\xi$.  Below $p_c$, $\rho_s$ vanishes as $|p-p_c|^x$,
$x \approx 1.31$.}
\label{figthree}
\end{figure}

The remainder of this paper discusses observable consequences of the picture suggested by the above model: that there is a stiff or ``renormalized classical'' phase where quantum effects modify the system quantitatively (by reducing the stiffness) but not qualitatively.
Such a model has been previously used to fit $T=0$ numerics at $p_c$~\cite{sandvik}, and similar renormalized classical physics applies to the undiluted 2D AF~\cite{chakravarty}.  Here we extend the renormalized classical picture to $T>0$ and $p<p_c$ (Fig.~\ref{figthree}).

At $p_c$, the correlation length $\lambda$ diverges as a power-law as $T\rightarrow 0$ rather than as an exponential, and the power-law is given by the classical resistivity exponent: $\lambda \sim ({\tilde J}/kT)^{\nu/t}$.   The classical Heisenberg model (i.e., ${\bf s}$ a classical unit vector in
~(\ref{dhaf})) has a crossover from threshold behavior to quasi-2D
behavior for $p<p_c$.

We can estimate this crossover using a one-step real-space RG.  The length scale that characterizes the crossover from threshold to 2D physics is the percolation length $\xi \sim |p-p_c|^{-\nu}$.  Estimating the coupling strength at this length from the threshold stiffness, and using this coupling strength and length scale in the standard expression for $\lambda$ in the renormalized classical 2D AF, gives for the correlation length
\beq
\lambda \sim \cases{a ({\tilde J}/kT)^{\nu/t} &if $({\tilde
J}/kT)^{\nu/t} \ll \xi/a$\cr \xi e^{2 \pi \rho_s/kT}&if $({\tilde J}/kT)^{\nu/t} \gg \xi/a$}
\label{corr}
\eeq
which is shown in Fig.~\ref{figthree}.  Here, $\nu=4/3$ and $\rho_s(p)\approx\tilde{J}|p-p_c|^x$, with $x = t \approx 1.31$.  This estimated value for $x$, which in our one-step RG controls the effective coupling in the 2D AF as $p \rightarrow p_c$, has previously been conjectured for the bulk modulus as $p\rightarrow p_c$~\cite{degennes}.

There are several sets of experimental and numerical data to which the above forms can be compared.  The $T=0$, $p=p_c$ numerics of Sandvik do show classical percolation scaling, which is partly explained by our construction of a model stable to quantum fluctuations in the links.  Correlation length measurements on the DHAF from neutron scattering~\cite{vajk} seem to
exhibit a crossover with dilution.  The crossover was
fit to~\cite{castroneto}
\beq \frac{\lambda}{a}= {e \over 8} {c/a \over 2 \pi \rho_s} {e^{2
\pi \rho_s / T} \over 1 + (4 \pi \rho_s/T)^{-\nu_T}}.
\label{clean} \eeq
At high-$T$, (\ref{clean}) like the classical model, follows a
power-law rather than an exponential.  An intermediate scaling
regime controlled by a bilayer multicritical
point~\cite{bilayer} has been proposed~\cite{vajk} to explain
deviation of observed scaling ($\nu_T \approx 0.7$) from the classical value
($\nu_T \approx 1.02$).


In conclusion, superconducting or antiferromagnetic order can exist to $p_c$ in 2D but has low stiffness because of both cluster geometry and quantum fluctuations.  This reduced $\rho_s$ results in a correlation length much shorter for $T>0$ than without dilution.  For diluted La$_2$CuO$_4$, our analytic results explain and compare reasonably well with existing experiment and numerics.

The authors acknowledge helpful conversations with M.\ Greven,
D.-H.\ Lee, T.\ Senthil, O.\ Starykh, O.\ Vajk, and J.\ Wu and
support from DOE LDRD-366464, NSF DMR-0238760, and NERSC.

\bibliographystyle{unsrt}
\bibliography{randmag}

\begin{thebibliography}{10}

\bibitem{ahn}
C.~H. Ahn, S.~Gariglio, P.~Paruch, T.~Tybell, L.~Antognazza, and J.~M.
  Triscone.
\newblock {\em Science}, 284:1152, 1999.

\bibitem{senthil}
T.~Senthil and S.~Sachdev.
\newblock {\em Phys. Rev. Lett.}, 77:5292, 1996.

\bibitem{coniglio}
A.~Coniglio.
\newblock {\em J. Phys. A}, 15:3829, 1982.

\bibitem{deutscher}
G.~Deutscher, I.~Grave, and S.~Alexander.
\newblock {\em Phys. Rev. Lett.}, 48:1497, 1982.

\bibitem{bunde}
S.~Havlin and A.~Bunde.
\newblock In A.~Bunde and S.~Havlin, editors, {\em Fractals and disordered
  systems}. Springer-Verlag, 1991.

\bibitem{eggert}
S.~Eggert, I.~Affleck, and M.~D.~P. Horton.
\newblock {\em Phys. Rev. Lett.}, 89:047202, 2002.

\bibitem{vajk}
O.~P. Vajk, P.~K. Mang, M.~Greven, P.~M. Gehring, and J.~W. Lynn.
\newblock {\em Science}, 295:1691, 2002.

\bibitem{sandvik}
A.~W. Sandvik.
\newblock {\em Phys. Rev. B}, 66:024418, 2002.

\bibitem{todo}
S.~Todo, H.~Takayama, and N.~Kawashima.
\newblock {\em Phys. Rev. Lett.}, 86:3210, 2001.

\bibitem{castroneto}
Y.C. Chen and A.~H.~Castro Neto.
\newblock {\em Phys. Rev. B}, 61:R3772, 2000.

\bibitem{bilayer}
O.~P. Vajk and M.~Greven.
\newblock {\em Phys. Rev. Lett.}, 89:177202, 2002.

\bibitem{lang}
K.~M. Lang, V.~Madhavan, J.~E. Hoffman, E.~W. Hudson, H.~Eisaki, S.~Uchida, and
  J.~C. Davis.
\newblock {\em Nature}, 415:412, 2002.

\bibitem{sachdev}
S.~Sachdev.
\newblock {\em Quantum phase transitions}.
\newblock Cambridge, 2000.

\bibitem{laughlin}
R.~B. Laughlin.
\newblock cond-mat/0209269.

\bibitem{trivedi}
A.~Ghosal, M.~Randeria, and N.~Trivedi.
\newblock {\em Phys. Rev. B}, 63:020505, 2001.

\bibitem{chakravarty}
S.~Chakravarty, B.~I. Halperin, and D.~R. Nelson.
\newblock {\em Phys. Rev. B}, 39:2344, 1989.

\bibitem{lanczos}
K.~Wu and H.~Simon.
\newblock Thick-restart {Lanczos} method for large symmetric eigenvalue
  problems.
\newblock {\em SIAM J. Matrix Anal. Appl.}, 22:602--616, 2000.

\bibitem{cardy}
J.~L. Cardy.
\newblock {\em Scaling and renormalization in statistical physics}.
\newblock Cambridge, 1996.

\bibitem{herrmann}
H.~J. Herrmann and H.~E. Stanley.
\newblock {\em Phys. Rev. Lett.}, 53:1121, 1984.

\bibitem{straley}
J.~P. Straley.
\newblock {\em J. Phys. C}, 15:2333, 1982.

\bibitem{degennes}
P.~G. de~Gennes.
\newblock {\em J. Phys. Lett.}, 37:L1, 1976.

\end{thebibliography}
\end{document}